\documentclass[twocolumn,superscriptaddress]{revtex4}
\usepackage{graphicx}
\usepackage{bm}
\usepackage{color}
\usepackage{amsmath}
\usepackage{amsfonts}
\usepackage{amssymb}
\usepackage{natbib}
\usepackage{latexsym}
\usepackage{mathrsfs}

\begin{document}

\title {Construction of New Solutions to Field Equations\\ by Using one Non--Separable Solution and one Symmetry of the System}

\author{Sergio A. Hojman}
\email{sergio.hojman@uai.cl}
\affiliation{Departamento de Ciencias, Facultad de Artes Liberales,
Universidad Adolfo Ib\'a\~nez, Santiago 7491169, Chile.}
\affiliation{Departamento de F\'{\i}sica, Facultad de Ciencias, Universidad de Chile,
Santiago 7800003, Chile.}
\affiliation{Centro de Recursos Educativos Avanzados,
CREA, Santiago 7500018, Chile.}

\begin{abstract}
Symmetries of the field equations are used to construct infinitely many non--trivial linearly independent new solutions to different partial differential equations such as the Schr\"odinger, the diffusion, and the paraxial equations, among many others, including Klein--Gordon, Dirac, Maxwell, Rarita--Schwinger and linear Einstein field equations (and even some especial seed solutions of fully non-linear General Relativity). The construction is done by applying one symmetry operator of the differential system to one non--separable seed solution of the same system.
\end{abstract}

\maketitle
 
\section{Introduction}

Symmetries are usually combined with the knowledge of  Lagrangian or Hamiltonian formulations of the evolution of (mechanical or field theoretical) dynamical systems to produce conservation laws to help integrating the equations of motion under consideration. The solutions to the differential system may also be achieved using other methods \cite{sah}.\\

In the present article I start from the knowledge of one symmetry, in conjunction with one known non--separable (seed) solution of a field theoretical dynamical system, to construct new non--trivial solutions for the same system. In general, some of the new solutions generated using this approach might not be linearly independent from the seed, in which case they are not of much use. These cases are not considered in this work.\\

This approach is based on related work \cite{sah,sah1}, where non--Noetherian symmetries of the equations of motion, which do not satisfy Noether's theorem, are used in the case of $n$--dimensional small oscillations problem to construct $n$ constants of motion in involution, which allow to devising a new method to solve that problem completely.\\

\section {The method}

In order to motivate the approach I will present it by using the familiar Schr\"odinger equation. The method is based in a general well--known result summarized by Eqs. \eqref{sch3} and \eqref{comm} below, and it is valid as long as they are fulfilled.\\

Nevertheless, in some of the very usual separable seed solutions cases, the procedure does not provide new non--trivial solutions as it is stated immediately before and after Eq. \eqref{sep}. The symmetry operator needs to be applied to a non--separable solution in order to obtain new non--trivial functions which solve the equation under consideration.\\

Let us start by considering the Schr\"odinger equation for an arbitrary time--independent potential $V=V(\vec{r})$

\begin{equation}\label{sch1}
\left(-\frac{\hbar^2}{2m} \vec{\nabla}^2 + V(\vec{r}) - i \hbar \frac{\partial}{\partial t} \right)  \psi(\vec{r},t)=0\ ,  
\end{equation}

\noindent and define the operator $\mathscr{H}$ by

\begin{equation}\label{sch2}
\mathscr{H} \equiv (H - i \hbar \frac{\partial}{\partial t})\ ,  
\end{equation}
to write the Schr\"odinger equation as
\begin{equation}\label{sch3}
\mathscr{H}\psi(\vec{r},t) = 0\ .
\end{equation}

It is well known that any symmetry operator $\mathscr{O}$ which commutes with $\mathscr{H}$ 

\begin{equation}\label{comm}
[\mathscr{H}, \mathscr{O}]  = 0\ ,
\end{equation}
\noindent defines a new solution to the Schr\"odinger equation $\psi_{\mathscr{O}}(\vec{r},t)=\mathscr{O} \psi(\vec{r},t)$ once a seed solution $\psi(\vec{r},t)$ is known.\\

We are presenting the method in terms of the Schr\"odinger equation but, of course, the explicit forms of the operators $\mathscr{H}$ and $\mathscr{O}$ are irrelevant as long as Eqs. \eqref{sch3} and \eqref{comm} are satisfied.\\

In the general setting defined by an arbitrary potential $V(\vec{r})$ as it appears in \eqref{sch1}, the only symmetry operator that can be constructed is the time--translation one, written either as $\mathscr{O}_H=H$ or, equivalently, as $\mathscr{O}_t=i \hbar \frac{\partial}{\partial t}$.\\

These two time--translation operators are equivalent when acting on solutions to the Schr\"odinger equation \eqref{sch1}.\\

The time--translation operator is the only (non--trivial) one that can be constructed because otherwise there will be a non--trivial universal symmetry operator defined for {\it {any}} potential $V(\vec{r})$.\\

The point is, that in spite of the fact that the above statement is correct, in most cases it is not of much use, because we deal usually with separable wave functions $\psi_s(\vec{r},t)$ of the form

\begin{equation}\label{sep}
\psi_s(\vec{r},t)\ = \ R(\vec{r})\ e^\frac{-i E t}{\hbar} \ , 
\end{equation}

\noindent and application of the time--translation operator to a solution $\psi_s(\vec{r},t)$ of Eq.\eqref{sch1} yields simply $E\ \psi_s(\vec{r},t)$, which is, of course, also a solution to Eq.\eqref{sch1}, but it {\it {is not a new solution}}.\\

Note that the mechanism presented in this work is useful for any time--independent (or any other kind of symmetric) $\mathscr{H}$ regardless of its detailed structure.\\

A few comments seem in order.\\

Even though is a known fact, it is important to recall that Feynman propagators may be used as wave functions to solve Schr\"odinger's equation for a time--independent potential using one of the end points (let's say the final point) as the space variable(s) and taking the initial point variable(s) as (dummy) parameters.\\

I should also mention that the same method may be used for field equations written on time--independent curved backgrounds or for backgrounds with other type of symmetries.\\ 

\section{The Bohm potential} 
Before applying the method to construct new solutions to field equations it seems appropriate to make a slight detour to introduce some facts and discuss the relevance of the existence of non--vanishing Bohm potentials in some of the examples I discuss in the next Sections.\\

\begin{enumerate}

\item There are quantum solutions describing the propagation of particles subject to different kinds of external potentials which exhibit unexpected behavior that disagrees with its classical one \cite{berry,ha201,ha20,ha211,hams21,ahms21}.\\ 

\item These surprising types of propagation have been experimentally confirmed \cite{sivi,bloch}. \\

\item This unusual results are due to the fact that the quantum solutions have non--vanishing Bohm potential \cite{ha201,ha20,ha211,hams21,ahms21}.\\

\end{enumerate}
In fact, classically, the external potential $V(\vec{r},t)$ determines completely the acceleration of a particle of mass $m$, while in quantum mechanics, the Bohm potential $V_{Bohm}(\vec{r},t)$, which depends on the functional form of the wave function, modifies the classical acceleration. The ``quantum'' acceleration $\vec{a_q}(\vec{r},t)$ may be computed as the sum of the classical acceleration  $\vec{a_c}(\vec{r})$ plus the Bohm acceleration $\vec{a_{Bohm}}(\vec{r},t)$       

\begin{eqnarray}\label{acc}
\vec{a_q}(\vec{r},t)&=&-\frac{1}{m}  \vec{\nabla}(V(\vec{r}) + V_{Bohm}(\vec{r},t)) \nonumber \\
&\equiv& \vec{a_c}(\vec{r})+\vec{a}_{Bohm}(\vec{r},t) \ .
\end{eqnarray}\\
This result agrees with the one obtained for the velocity computed as $\vec{p}/m= \vec{\nabla} S/m$ \cite{ha20}.\\

Consider one wavefunction  $\psi=\psi(\vec{r},t)$, that solves the Schr\"odinger equation \eqref{sch1} and its complex conjugate counterpart.\\

The polar version of the wavefunction \cite{mad,bohm,holland,wyatt,ha20}, may be written in terms of $A(\vec{r},t)$ and $S(\vec{r},t)$ that are real functions of space and of time as
\begin{equation}\label{psi}
{{\psi}}(\vec{r},t) = A(\vec{r},t)\ e^{i S(\vec{r},t)/\hbar}\, .     \end{equation}
Now, write the real and imaginary parts of the Schr\"odinger equation respectively as,
\begin{eqnarray}
\frac{1}{2m} ({\vec{\nabla}S})^2 - \frac{\hbar^2}{2m}\frac{  \vec{\nabla}^2 A }{A} + V +\frac{\partial S}{\partial t} &=&0\, , \label{HJB1} \\
\frac{1}{m}\vec{\nabla}\left(A^2 {\vec{\nabla}S}\right)  +\frac{\partial A^2}{\partial t}&=&0\, . \label{cont}    
\end{eqnarray}

The first equation \eqref{HJB1} is sometimes called the quantum Hamilton--Jacobi equation for the external potential $V(\vec{r})$, which is a modified version of its classical counterpart \cite{holland,wyatt}. The second equation \eqref{cont} is the probability conservation equation. The classical Hamilton--Jacobi equation appears modified by the addition of the Bohm potential $V_{Bohm}(\vec{r},t)$ defined by
\begin{equation}
V_{Bohm} (\vec{r},t) \equiv  - \frac{\hbar^2}{2m}\frac{  \vec{\nabla}^2 A }{A}\, . \label{VB2}    
\end{equation}
In general, the Bohm potential does not vanish. There is a family of external potentials \cite{ha20} which admit some solutions for which the Bohm potential vanishes, but even for those potentials, most of their solutions produce non--vanishing Bohm potentials. Most of the potentials have only non--vanishing Bohm potential solutions.\\

 Even the free particle, which admits planes waves with vanishing Bohm potentials as solutions, has solutions such as \eqref{airy1} and \eqref{airy2} with non--vanishing Bohm potentials, which exhibit accelerations given by \eqref{acc} that do not vanish, in general, even for the case $V(\vec{r})=0$.\\

\section{Application of the method to non--separable seed solutions}

I will start by applying this method to the diffusion equation.\\ 

\subsection{The diffusion equation} 

Consider the diffusion equation

\begin{equation}\label{diff}
{\vec{\nabla}}^2 u(\vec{r},t)-\frac{\partial u(\vec{r},t) }{\partial t} = 0\ ,
\end{equation}
 
\noindent which may be solved by the (non--separable) heat kernel $U_{HK}(\vec{r}_1,\vec{r}_2,t)$ 

\begin{equation}\label{hk}
U_{HK}(\vec{r}_1,\vec{r}_2,t)=\frac{e^{-(\vec{r}_2-\vec{r}_1)^2/4 t}}{\left(4\pi t\right)^{3/2}} \ .
\end{equation}

Compute the time derivative of the heat kernel, $\mathscr {O}_t U_{HK}(\vec{r}_1,\vec{r}_2,t)/{i \hbar}$ to get

\begin{equation}\label{Ut}
\frac{\partial U_{HK}(\vec{r}_1,\vec{r}_2,t)}{\partial t} =\frac{-6t^2+(\vec{r}_2-\vec{r}_1)^2}{4 t^2}\ U_{HK}(\vec{r}_1,\vec{r}_2,t)\ ,
\end{equation}
\noindent which also solves the diffusion equation.\\

The second time derivative of the heat kernel
\begin{eqnarray}\label{Utt}
\frac{\partial^2 U_{HK}(\vec{r}_1,\vec{r}_2,t)}{\partial t^2} &=&\frac{60 t^2 - 
 20 t (\vec{r}_2-\vec{r}_1)^2 + ((\vec{r}_2-\vec{r}_1)^2)^2}{16 t^4} \nonumber \\& & U_{HK}(\vec{r}_1,\vec{r}_2,t)\ ,
\end{eqnarray}
\noindent is also a solution to the diffusion equation, so on and so forth.\\

The same method may be applied to derivatives with respect to the space variables because equation \eqref{diff} is also space--translation invariant.\\

Consider the derivative of the heat kernel with respect to $x_1$, $\partial U_{HK}(\vec{r}_1,\vec{r}_2,t)/\partial x_1$, for instance, to get

\begin{equation}\label{Ux1}
\frac{\partial U_{HK}(\vec{r}_1,\vec{r}_2,t)}{\partial x_1} =\frac{(x2 - x1)}{2 t}\ U_{HK}(\vec{r}_1,\vec{r}_2,t)\ ,
\end{equation}
\noindent which also solves the diffusion equation.\\

By essentially the same token, integration of the solutions also works. Take, for instance,

\begin{equation}\label{Uintx1}
\int U_{HK}(\vec{r}_1,\vec{r}_2,t) dt = \ -\ \frac{\text{erf}(\frac{\sqrt{(\vec{r}_2-\vec{r}_1)^2}}{2 \sqrt{t}})}{{ 4 \pi \sqrt{(\vec{r}_2-\vec{r}_1)^2}}}\ ,
\end{equation}
which also solves the diffusion equation.\\

The same procedure may be iterated with mixed derivatives and integrals to get infinitely many solutions to the diffusion equation. The same strategy may be used again in the examples presented below.\\

Note that the time derivative of the heat kernel \eqref{Ut} is a again a (new and different) heat kernel which produces the evolution the initial condition $u(\vec{r}_1,\vec{r}_2,0)$ to $\partial u((\vec{r}_1,\vec{r}_2,t)/ \partial t$ which is, of course, a (new and different) solution to the diffusion equation. A similar outcome is obtained for the Feynman propagators in the Schr\"odinger and paraxial equations.\\

\subsection{The Schr\"odinger equation for free particles} 

Let us turn our attention to non--separable solutions to the {\it{free}} one--dimensional Schr\"odinger equation, such as the one found by Berry and Balazs in 1979 \cite{berry}, in terms of the Airy function denoted $Ai$,

\begin{eqnarray}\label{airy1}
\psi_{Airy}(x,t)&=&{Ai} \left(\frac{B}{\hbar^{2/3}} \left(x-\frac{B^3}{4m^2} t^2 \right)\right) \nonumber\\
& &e^{\frac {i B^3 t\left (6 m^2 x-B^3 t^2 \right)} {12 m^3 \hbar}}\ \ ,
\end{eqnarray}
\noindent which has remarkable properties.\\

This solution describes the motion of {\it{free}} propagation of waves (constant refractive index in the case of the paraxial equation, which is formally identical to the free Schr\"odinger equation) which {\it{exhibits constant acceleration}}. This unexpected fact has been observed experimentally both using electromagnetic waves in 2007 \cite{sivi} and electrons in 2013 \cite{bloch}.\\

In this case, the application of the time--translation operator to solution \eqref{airy1} 
\begin{eqnarray}\label{airy2}
\mathscr{O}_t \psi_{Airy}&=&\frac {-i B^3 e^{\frac {i B^3 t\left (6 m^2 x-B^3 t^2 \right)} {12 m^3\hbar}}}{ {4 m^3}} \nonumber \\ 
         &\Bigg[& 2 B {\hbar}^{1/3}  m t\ {Ai}\ ' \left(\frac {B\left(x - \frac {B^3 t^2} {4 m^2} \right)} {\hbar^{2/3}}\right) \nonumber \\ 
       &+& i\left (B^3 t^2 - 2 m^2 x \right){Ai}\left (\frac {B\left (x - \frac {B^3 t^2} {4 m^2} \right)} {\hbar^{2/3}} \right) \Bigg]\ , \nonumber \\ 
\end{eqnarray}
\noindent is far from trivial.\\

This new solution to the {\it{free}} Schr\"odinger equation exhibits {\it{$x$-- and $t$--dependent acceleration}}.\\

Consider now the Feynman propagator for the free particle $K_{FP}(\vec{r}_1,\vec{r}_2,t)$

\begin{equation}\label{kfp}
K_{FP}(\vec{r}_1,\vec{r}_2,t)=\frac{m^{3/2}\ e^{i m(\vec{r}_2-\vec{r}_1)^2/2 \hbar i t}}{\left(2\pi \hbar i t\right)^{3/2}} \ ,
\end{equation}

\noindent which solves the free particle Schr\"odinger equation and it is extremely similar to the heat kernel. All what was done in the preceding subsection for the heat kernel $U_{HK}(\vec{r}_1,\vec{r}_2,t)$ may be repeated, {\it{mutatis mutandi}}, for the Feynman propagator $K_{FP}(\vec{r}_1,\vec{r}_2,t)$ here.\\

The same applies to all the other known propagators (considered as solutions to the Schr\"odinger equation as functions of ${\vec{r}_2}$ and time, for fixed ${\vec{r}_1}$) for time--independent potentials such as the harmonic oscillator.\\

\subsection{The paraxial equation} 

The paraxial equation for wave optics is written for the complex wave amplitude $A(x,y;z)$

\begin{equation}\label{par1}
\left(\frac{1}{2k^2}\left( \frac{\partial^2 }{\partial x^2} +\frac{\partial^2 }{\partial y^2}\right) - V(x,y) +\frac{i}{k} \frac{\partial }{\partial z} \right) A(x,y;z) =0\ .  
\end{equation}
This equation is similar to both the Schr\"odinger equation and the diffusion equation, so the method may be applied to it directly, taking into account that the refractive index $n(x,y)$ and the potential $V(x,y)$ are related by

\begin{equation}\label{nV}
V(x,y)=-\frac{n(x,y)-n_0}{2 n_0}\ ,
\end{equation}

\noindent where $k$ is a constant, $n_0$ is the constant refractive index of the background medium (see \cite{marte}, for details).\\

\subsection{The Schr\"odinger equation for the harmonic oscillator}

The method may, of course, also be applied to the Schr\"odinger equation beyond the free particle. Consider the one--dimensional harmonic oscillator and the (non--separable) Feynman propagator (or seed solution)

\begin{eqnarray}\label{kho}
K_{HO}(x_1,x_2,t)&=&\sqrt{\frac{m \omega}{2 \pi i \hbar \sin(\omega t)}} \nonumber \\
&\times&{\text{Exp}}\left(-\frac{ m \omega (({x_1}^2 + {x_2}^2) \cos(\omega t )}{2 i \hbar \sin(\omega t)}\right) \nonumber \\ 
&\times&{\text{Exp}}\left(\frac{2 x_1  x_2}{2 i \hbar \sin(\omega t)}\right), \nonumber \\
\end{eqnarray}
which satisfies
\begin{equation}\label{schho}
\left(-\frac{\hbar^2}{2m} \frac{\partial^2}{\partial {x_2}^2} +\frac{1}{2}m\omega^2 {x_2}^2 - i \hbar \frac{\partial}{\partial t} \right)  K_{HO}(x_1,x_2,t)=0\ , 
\end{equation}

\noindent or,
\begin{equation}\label{schho}
\mathscr {H}_{HO}  K_{HO}(x_1,x_2,t)=0\ . 
\end{equation}
It is a straightforward matter to show that, due to the $t$--traslation invariance of  Eq. \eqref{schho}, ${K^n}_{HO}(x_1,x_2,t)$ defined by

\begin{equation}\label{knho}
{K^n}_{HO}(x_1,x_2,t)=\frac{\partial^n}{\partial t^n}K_{HO}(x_1,x_2,t) \ . 
\end{equation}
\noindent also solves Eq. \eqref{schho} for all integer $n$.\\

Compute now ${K^1}_{HO}(x_1,x_2,t)$

\begin{eqnarray}\label{k1ho}
{K^1}_{HO}(x_1,x_2,t)&=&\frac{\partial}{\partial t}K_{HO}(x_1,x_2,t) \nonumber \\
&=& \bigg( \frac{(-2 i m {\omega}^2 ({x_1}^2 + {x_2}^2 -2 x_1 x_2 \cos(\omega t))}{4 \hbar\sin^2(\omega t)} \nonumber \\
& &-\frac{\hbar \omega \sin(2 \omega t)}{4 \hbar \sin^2(\omega t)} \bigg) \times {K}_{HO}(x_1,x_2,t)\ , 
\end{eqnarray}
\noindent which solves Eq. \eqref{schho}.\\

The method may be applied to any Hamiltonian $\mathscr{H}$ for which a non--separable seed is known. In the case of all the $x$--dependent Hamiltonians, the $x$--traslation is not a symmetry operation.\\

It is worth noting that the harmonic oscillator  $a^\pm$ ladder operators
\begin{equation}\label{ladder}
 a^\pm=\sqrt{\frac{ m \omega}{2\hbar}}x_2 \mp 
 \sqrt{\frac{\hbar}{2m \omega}}\ \frac{\partial}{\partial x_2}
 \end{equation}
have the usual commutation relations with $H_{HO}$ (and $\mathscr {H}_{HO}$)
\begin{equation}\label{Ha}
[\mathscr {H}_{HO}, a^\pm]=\  \pm\ \hbar \omega\ a^\pm\ ,
 \end{equation}
\noindent or
\begin{equation}\label{H-a}
(\mathscr {H}_{HO}\mp\hbar \omega) a^\pm K_{HO}(x_1,x_2,t)=\ 0\ ,
\end{equation}
which means that the application of the ladder operators to the Feynman propagator of the harmonic oscillator allows us to produce wave functions which solve the Schr\"odinger equation of a ``shifted'' Hamiltonian $(\mathscr {H}_{HO}\mp\hbar \omega)$. Of course, this operation may be iterated and may be used in conjunction with the $t$--traslation operation because it commutes with the ladder operators.\\

\section{Applications to solutions to massless field equations of any spin}
Consider $4$ orthogonal pre--potentials $u^{(\alpha)}(x^\beta)$, with greek indices ranging from $1$ to $4$, defined by \cite{ha212}
\begin{eqnarray}\label{dal}
 \square  u^{(\alpha)} &=& 0\ , \ \ \forall \alpha \, ,\nonumber\\
{{u^{(2i-1)}}_{,\alpha}} {u^{(2i)}}^{,\alpha} &=& 0\ , 
\end{eqnarray}
for  $i=1, 2$ and where $\square$ is the d'Alembert operator in Minkowski space.\\

It has been proved that the pre--potentials so defined may be used to generate solutions to any massless field equations for any spin, i.e., Klein--Gordon, Dirac, Maxwell, Rarita--Schwinger and linearized Einstein equations (for details, see \cite{ha212}).\\

In some instances, the same pre--potentials also solve the full non--linear Einstein equations. I will present one such example of the application of the method presented in this paper to pre--potentials thus applicable to massless equations of any spin, including General Relativity.\\

Consider the following pre--potentials

\begin{eqnarray}\label{prepo}
u^{(1)}(x-t)&=&\tanh(x-t)\, ,\nonumber\\
u^{(2)}(y,z)&=&y^3-3yz^2\, ,\nonumber\\
u^{(3)}(x,y,z,t)&=&0\, ,\nonumber\\
u^{(4)}(x,y,z,t)&=&0\, .
\end{eqnarray}
Let us first state that
\begin{equation}
\phi(x,y,z,t)=u^{(1)}(x-t) \times u^{(2)}(y,z)  
\end{equation}
solves the massless Klein--Gordon equation. Of course, $u^{(1)}(x-t)$ and $u^{(2)}(x-t)$ solve it too, but we are now interested in non--linear functions of the pre--potentials and their derivatives only.\\

The antisymmetric $F_{\alpha \beta}$ field
\begin{eqnarray}\label{F4}
F_{\alpha \beta}(x^\gamma)&=&{u^{(1)}}_{,\alpha} {u^{(2)}}_{,\beta}-{u^{(2)}}_{,\alpha} {u^{(1)}}_{,\beta}\nonumber \\
&+&{u^{(3)}}_{,\alpha} {u^{(4)}}_{,\beta}-{u^{(4)}}_{,\alpha} {u^{(3)}}_{,\beta}\ ,
\end{eqnarray}
satisfies all of the (source free) Maxwell equations.\\

For Einstein equations, define the metric as
\begin{equation}
g_{\alpha\beta}={\hat g}_{\alpha\beta}+\epsilon\ \Theta_{\alpha\beta}\, ,
\end{equation}
\noindent where ${\hat g}_{\alpha\beta}={\text{diag}}(-1,+1,+1,+1)$ , $\epsilon<<1$ for linearized gravity,  $\epsilon=1$ for fully non--linear General Relativity and 

\begin{eqnarray}
\Theta_{\alpha \beta}(x^\gamma)&=&{{u}^{(1)}}_{,\alpha} { {u}^{(2)}}_{,\beta}+{ {u}^{(2)}}_{,\alpha} { {u}^{(1)}}_{,\beta}\nonumber\\ &+&{{u}^{(3)}}_{,\alpha} { {u}^{(4)}}_{,\beta}+{ {u}^{(4)}}_{,\alpha} { {u}^{(3)}}_{,\beta}. 
\end{eqnarray}
The metric $g_{\alpha\beta}$ solves both linearized and full Einstein's equations and gives rise to a non--vanishing Riemann tensor (even at order $\epsilon$ for linearized gravity).\\

To solve the massless Dirac equation, it is enough to define $\psi (x^\alpha)$ by
\begin{equation}
\psi (x^\alpha)  = \gamma^{\rho} {\partial}_{\rho} \begin{pmatrix}
 u^{(1)}(x^\mu)\ u^{(2)}(x^\nu)\\
 u^{(3)}(x^\mu)\ u^{(4)}(x^\nu)\\
 u^{(5)}(x^\mu)\ u^{(6)}(x^\nu)\\
 u^{(7)}(x^\mu)\ u^{(8)}(x^\nu)\\
\end{pmatrix} \ ,
\end{equation} 

\noindent where one may take all the, up to now, undefined pre--potentials to be zero (for more interesting solutions, see \cite{ha212}).\\

To solve the massless Rarita--Schwinger equation one may take the vector--spinor $\psi_\beta (x^\alpha)$ given by
\begin{equation}
\psi_\beta (x^\alpha)  ={\partial}_\beta\left(\gamma^{\rho} {\partial}_{\rho} u(x^\mu)\right) \gamma^{\tau} {\partial}_{\tau}\begin{pmatrix}
 u^{(1)}(x^\nu)\\
 u^{(2)}(x^\nu)\\
 u^{(3)}(x^\nu)\\
 u^{(4)}(x^\nu)\\
\end{pmatrix} \ .
\end{equation} 
Again, more interesting solutions are presented in \cite{ha212}.\\

It is remarkable that {\it{the same pre--potentials}} \eqref{prepo} may be used {\it{to solve all of the masless field equations for any spin, including fully non--linear General Relativity}}.\\

Furthermore, the partial derivatives of any order of $u^{(1)}(x-t)$ with respect to $x$ or with respect to $t$ (as well as mixed derivatives) and the partial derivatives of any order of $u^{(2)}(y,z)$ with respect to $y$ or with respect to $z$ (as well as mixed derivatives) give rise to new pre--potentials (which satisfy \eqref{dal}) and new non--trivial solutions for the massless field equations of any spin including (non--linearized) General Relativity.\\ 

\section{Summary}

I have presented a non--trivial application of symmetry operators to generate infinitely many new and non--trivial solutions (or Feynman propagators or kernels) to any equation which has at least one non--trivial symmetry and  for which a non--separable seed solution (or Feynman propagator or kernel) is known.\\

I have applied the method to generate non--trivial and new (as far as I know) solutions to many of the most familiar field equations used in physics including fully non--linear General Relativity.


\begin{thebibliography}{}
\bibitem{sah} Sergio A. Hojman, J. Math. Phys. {\bf 34}, 2968 (1993).
\bibitem{sah1} Sergio A. Hojman, {\it {``A New Way to Solve the Small Oscillations Problem''}}, (2021).
\bibitem{berry} M. V. Berry and N. L. Balazs, Am. J. Phys. {\bf 47}, 264 (1979).
\bibitem{ha201} S. A. Hojman and F. A. Asenjo,  Phys. Lett. A {\bf 384}, 126263 (2020).
\bibitem{ha20} S. A. Hojman and F. A. Asenjo, Phys. Scr. {\bf 95}, 085001 (2020).
\bibitem{ha211} F. A. Asenjo and S. A. Hojman, Eur. Phys. J. C {\bf 81}, 98 (2021).
\bibitem{hams21} S. A. Hojman, F. A. Asenjo, H. M. Moya--Cessa and F. Soto--Eguibar, Optik {\bf 232}, 166341 (2021) 
\bibitem{ahms21} F. A. Asenjo, S. A. Hojman,  H. M. Moya--Cessa and F. Soto--Eguibar, Optics Communications {\bf 490} 126947 (2021)

\bibitem{sivi} G. A. Siviloglou, J. Broky, A. Dogariu and D. N. Christodoulides, Phys. Rev. Lett. 99, 213901 (2007).
\bibitem{bloch} N.  Voloch-Bloch, Y. Lereah, Y. Lilach, A. Gover and A. Arie, Nature {\bf 494}, 331 (2013).

\bibitem{mad} E. Madelung, Zeit. f. Physik {\bf 40}, 322 (1927).
\bibitem{bohm} D. Bohm, Phys. Rev. {\bf 85}, 166 (1952).
\bibitem{holland} P. R. Holland, {\it The Quantum Theory of Motion: an account of the de Broglie-Bohm causal interpretation of quantum mechanics}, (Cambridge University Press, 1993).
\bibitem{wyatt} R. E. Wyatt, {\it Quantum Dynamics with Trajectories: introduction to quantum hydrodynamics} (Springer, 2005).

\bibitem{marte} M.A. Marte and S. Stenholm, Phys. Rev. A {\bf 56}, 2940 (1997).

\bibitem{ha212} S. A. Hojman and F.A. Asenjo, {\it{``Unification of massless field equations solutions for any spin''}}, arXiv:2102.01485, quant-ph.

\end{thebibliography}
\end{document}